# Size Matters: The Use and Misuse of Statistical Significance in Discrete Choice Models in the Transportation Academic Literature

**Giancarlos Parady**
Lecturer
Department of Urban Engineering
The University of Tokyo, Bunkyo-ku, Tokyo-to, Japan, 113-8656
Email: gtroncoso@ut.t.u-tokyo.ac.jp
ORCiD: 0000-0002-7581-774X

**Kay W. Axhausen**
Professor
IVT, ETH Zürich, CH-8093 Zürich
E-mail: axhausen@ivt.baug.ethz.ch

**ABSTRACT**
In this paper we review the academic transportation literature published between 2014 and 2018 to evaluate where the field stands regarding the use and misuse of statistical significance in empirical analysis, with a focus on discrete choice models. Our results show that 39% of studies explained model results exclusively based on the sign of the coefficient, 67% of studies did not distinguish statistical significance from economic, policy or scientific significance in their conclusions, and none of the reviewed studies considered the statistical power of the tests. Based on these results we put forth a set of recommendations aimed at shifting the focus away from statistical significance towards proper and comprehensive assessment of effect magnitudes and other policy relevant quantities.
**Keywords:** Discrete choice models, effect size, statistical significance, statistical power, policy and practice implications

# 1. INTRODUCTION

Generally speaking, the purpose of academic transportation research is to better understand transport-related human behaviour to better inform transportation policy design and implementation (*1*), and with the advent of cheap computing power, statistical models have become a key tool to help explain transport-related phenomena. However, along with the widespread use of statistical models across many fields relying on quantitative analysis, came its widespread "cookbook" use, where statistical significance is used to define the importance, or lack thereof, of any given variable, regardless of its practical importance. The seminal work of McCloskey and Ziliak (*2*) shed some light on consistent errors in the use of "statistical significance" in the field of economics. But this is not by any means exclusive to the economics field, and while we suspect the transportation field is no stranger to such errors, this has yet to be systematically evaluated, which motivates this work. In this paper we adapt McCloskey and Ziliak (1996)'s 19 questions to the academic transportation literature to evaluate where the field stands regarding the use and misuse of statistical significance in empirical analyses, with a focus on discrete choice models. This study complements the study of Parady, Ory & Walker (*1*) on model validation in an attempt to promote better modeling practices in the field.

# 2. McCLOSKEY AND ZILIAK'S KEY FINDINGS

As an unbiased selection of best practice in economics, McCloskey and Ziliak (*2*) reviewed all full-length papers published in the American Economic Review in the 1980s that used regression analysis. For the 182 papers selected, they asked 19 questions about the use of statistical significance, and recorded answers as "yes" for sound statistical practice, "no" for unsound practice, or "not applicable." Their key findings are summarized below:

1. 70% of reviewed studies did not distinguish statistical significance from economic, policy or scientific significance.
2. One third of studies used only the t- and F-statistics as criteria for variable inclusion in the analysis of the paper.
3. 72% of studies did not discuss the scientific conversation within which the magnitude of a coefficient can be judged to be "large" or "small."
4. 59% of studies ambiguously used the word "significant" to mean statistically different from the null sometimes and to mean practically important at other times.
5. 32% explicitly stated to have used statistical significance as an exclusive criterion to drop variables from a model.
6. Only 4% of studies considered the statistical power of their tests.
7. 69% of studies did not report descriptive statistics of variables used in the model.

# 3. EVALUATING THE USE OF STATISTICAL SIGNIFICANCE IN THE TRANSPORTATION ACADEMIC LITERATURE

## 3.1 Article selection criteria

To get a more comprehensive view of the current state of affairs, we extended the scope of the review to the whole field, rather than a "selection of best practice" in the field. We based our selection criteria on Parady, Ory and Walker (*1*) so that the results presented here can complement their findings on validation practices. A key difference, however, is that we did not exclude papers using stated preference surveys from the review. Using the Web of Science Core Collection maintained by Clarivate Analytics we reviewed discrete choice model reporting practices in the academic transportation literature published between 2014 and 2018. Articles were selected based on the following criteria:

1. Peer-reviewed journal articles published between 2014 and 2018
2. Analysis uses discrete choice models
3. Target choice dimensions are:
   a) Destination choice
   b) Mode choice
   c) Route choice

4. Articles that analyse other choice dimensions are considered if and only if the article includes at least one of the three target choice dimensions defined in 3.
5. Web of Science Database search keywords are:
    a) Destination choice model
    b) Mode choice model
    c) Route choice model
6. Web of Science Database fields are:
    a) Transportation
    b) Transportation science and technology
    c) Economics
    d) Civil engineering
7. Research scope is limited to human land transport and daily travel behaviour (tourism, evacuation behaviour, and freight transport articles were excluded)
8. Studies using numerical simulations only were excluded
9. Methodological papers are only included if they used empirical data. In addition, as a new criterion for this study, papers must have a clear focus on policy or policy related variables (papers whose stated contribution is exclusively methodological were excluded)
10. For route choice models, Stochastic user equilibrium (SUE) models are excluded, as discrete choice models are just a subcomponent of a larger model.

The final number of articles reviewed was 95 articles, selected randomly out of a total of 283 articles. This is equivalent to 34% of all articles matching the inclusion criteria.

### 3.2 The 15 questions for the transportation field
While this study is based on McCloskey and Ziliak (*2*), we have adapted the questions to reflect the idiosyncrasies of the transportation field. In addition, for some questions, we expanded the yes-no dichotomy to include partially satisfactory practices. While whenever possible we tried to align our scoring criteria with their criteria, the way some questions were operationalized was not clearly described in their paper, so it is possible that there are differences in scoring criteria and methods. We advise the reader to keep this in mind as we compare both studies in the next subsections.

To adapt the questions to the field, we started with an a-priori set of questions adapted based on the authors experience, and pre-tested them on a random subset of the eligible papers (N=29), iteratively modifying the questionnaire until reaching a satisfactory set of 15 questions. Moving forward, when referring to McCloskey and Ziliak's 19 questions we use the nomenclature MZ-1, … , MZ-19. The 15 questions for the transportation field are:

*Q1. Does the paper report descriptive statistics and units for model variables?*
This question is equivalent to MZ-2, and points to the fact that knowledge of variable units and basic descriptive statistics (at least, means for continuous variables and relative frequency for categorical variables) is crucial to properly interpret model results. We have extended the answers to "yes, largely," " yes, partially" and "not at all"

*Q2. Are estimated coefficients used to calculate elasticities, marginal effects or some other quantity of interest that addresses the question of "how large is large"?*
This is equivalent to MZ-3, but in discrete choice modeling this is even more critical given that, as opposed to linear regressions, coefficient estimates are not directly interpretable. In transportation, elasticities, marginal effects, and to a lesser extent odd ratios are usually reported. Other quantities of interest usually reported are marginal rates of substitution, in particular, the value of travel time. While such quantities are not strictly speaking measures of effect size, they are quantities of policy importance, hence within the scope of this and subsequent questions.
Note that amount of explained variance is often referred to as a measure of effect size. While we agree that this measure does convey an idea of a set of variables or a model's explanatory power, such measure does not have a direct policy interpretation in terms of magnitude in the way an elasticity or a marginal effect does. In that regard it is not relevant here and subsequent questions.
.
      We classified papers into three categories: "yes, in a comprehensive manner," for such cases where effects sizes or other quantities of interest are reported for (i) most variables in the paper or (ii) those variables the authors have explicitly identified as important in the objectives or hypotheses statements of the paper; "yes, partially," for the cases when these are only reported for one or a few variables only, but not necessarily covering all the key variables; and "not at all."
      Since reporting of coefficients is a convention of the field, this question ignores whether or not coefficients are reported in the first place. That is, a paper that reports measures of effect magnitudes for all variables in addition to the model coefficients, and a paper that exclusively reports measures of magnitude for all variables, will both be classified as "yes, in a comprehensive manner."

*Q3. Does the paper report all standard errors, t-statistics, and goodness of fit statistics like the likelihood ratio test, and rho-square?*
This question is an alternative to MZ-6, which asks if "*the paper eschews reporting all t- or F-statistics or standard errors, regardless of whether a significance test is appropriate.*" Since it is a convention in the field that null hypotheses are implied based on the objectives of the study, full model results including coefficients, t-statistics, and goodness of fit statistics are usually reported, and in our experience demanded by reviewers if not. We reformulated the question to evaluate to what extent the model reporting convention is met (coefficients, significance statistics, and goodness of fit), with the explicit understanding that coefficients and significance test statistics should not be given a primary position in a paper over the results directly related to effect magnitudes and policy discussions.

*Q4. Does the paper consider the power of the test?*
This is equivalent to MZ-8, referring to the statistical power of a test.

*Q5. If so, does it do anything about power?*
This is equivalent to MZ-9.

*Q6. Does the paper eschew "asterisk econometrics," the ranking of coefficients according to the absolute size of the test statistic?*
This is equivalent to MZ-10. This practice is not conventional in the field, so we would expect a priori to not happen very often.

*Q7. In the model results section, does the paper eschew "sign econometrics," remarking on the sign but not the magnitude of the effect?*
This is equivalent to MZ-11. It refers to the practice of describing models based on the sign of the coefficient (usually in addition to the size of the t-statistic) without considering the magnitude of the effect in question, and whether such effect is large enough to matter in practical terms. We explicitly limited the scope of this question to the section where the model results are first introduced to account for the fact that it is plausible a researcher completely discusses a model in terms of sign econometrics (a practice we discourage while acknowledging the role conventions play in perpetuating this) but then proceed to conduct some separate analysis that does give some idea of magnitude. This includes remarking on quantities derived from coefficients such as the value of travel time.
      We classified papers into three categories: "yes, comprehensively" for cases where "sign econometrics" are eschewed for most variables in the paper, "yes, partially" when this is the case for one or a few variables only, but not necessarily covering all the key variables; and "not at all."

*Q8. Does the paper discuss the magnitude of estimated effects or other quantities of interest?*

*Parady and Axhausen*

*Q9. Does the paper make a judgement on magnitudes, making the point that some effects or quantities of interest are practically influential or important and some are not?*
These questions relate to MZ-12 and focus on whether the paper makes the point of the practical importance (as opposed to statistical significance) of observed effects. While in their original questionnaire they refer specifically to coefficients, as we noted earlier, coefficients are not directly interpretable in discrete choice models, so we expanded the scope of this question to cover any form of analysis that focuses on magnitude, including simulations. Furthermore, in the pre-test phase we decided to split this question in two parts. Q8 refers to whether or not there was a discussion of magnitude, either by discussing elasticities, marginal effects, marginal rates of substitution, or by conducting simulation analyses. Such discussions include the interpretation of estimated magnitudes and their relative comparisons. Q9, on the other hand, explicitly asks whether the authors make a judgement on the magnitudes observed, that is, whether they explicitly judge an effect or quantity of interest to be "large," "medium," "small" or "practically important" or "practically negligible" based on some criteria. This distinction is important, because the largest effect among a group of effects might still be small in practical terms. And while it might seem trivial at first, such judgement of magnitudes is an important part of a quantitative study, and it is not necessarily that straightforward a task.

For these two questions, we classified papers into three categories: "yes, comprehensively" for such ideal cases where they do so for (i) most variables in the paper or (ii) those variables the authors have explicitly identified as important in the objectives or hypotheses statements of the paper; "yes, in a limited manner" when they do so for one or a few variables only, but not necessarily covering all the key variables, and "not at all."

*Q10. Does the paper discuss the scientific conversation within which an effect or other quantity of interest would be judged large or small?*
This is equivalent to MZ-13. It asks whether the author compares her own findings against previous studies in the literature or commonly accepted values in the field. We marked this question as "yes" if at least one quantity of interest is compared against the literature.

*Q11. Does the paper avoid choosing variables for inclusion solely on the basis of statistical significance?*
This is equivalent to MZ-14. It asks whether authors explicitly state dropping variables from a model based exclusively on statistical significance, disregarding the magnitude of the effect. Papers using stepwise variable selection methods are also marked as "no." As in Ziliak and Mcloskey (*2*) papers are only marked "no" when authors state so explicitly, so this can be thought of as a lower bound of this criterion.

*Q12. Does the paper do a simulation to determine whether the estimated effects or other quantities of interest are reasonable and/or to better illustrate the magnitude of estimated effects?*
This is equivalent to MZ-17. It also includes policy simulations. Note that this is different from the case where simulation is necessary to estimate magnitudes, for example marginal effect of dummy variables, or elasticities in open-form models, although we recognize in some instances this difference might be blurry.

*Q13. In the conclusion and implication sections, is statistical significance kept separate from economic policy and scientific significance?*
This is equivalent to MZ-18. For example, papers that conclude summarizing variables that were found to be statistically significant and even proceed to infer policy suggestions from these "findings" are marked as "no" as they are mixing statistical significance with practical importance. The scope of this question is limited to the part of the conclusion and implication sections that refer directly to model results.

*Q14. In the estimation, conclusion and implication sections, does the paper avoid using the word "significance" in ambiguous ways, meaning "statistically significant" in one sentence and "large enough to matter for policy or science" in another?*
This is equivalent to MZ-19. We limited the scope of this question to the estimation, conclusion, and implication sections.

*Q15. Does the article report confidence intervals of effect sizes, using them to interpret practical importance and not merely as a replacement for pointwise statistical significance?*
This is technically not part of the 19 original questions, but a question #20 McCloskey and Ziliak wished they could have added (*3*) . Note that in this question we refer to the confidence interval of a measure of effect magnitude such as elasticities or marginal effects, not of coefficients.
  We classified papers into three categories: "yes, in a comprehensive manner," "yes, in a limited manner," and "not reported for any variable." In this case, "yes, in a limited manner" refers to the case where confidence intervals are reported but not used in the discussion.

*Questions eliminated from the questionnaire*
  Readers will note that we have indeed eliminated some questions from the original questionnaire. MZ-1 refers to the use of a small number of observations such that statistically significant differences are not just a result of large sample size. To avoid the issue of what a "small number" of observations is, we excluded this question. Instead, we report the distribution of the minimum sample sizes used in the reviewed studies (see Section 4.2).
  MZ-4 asks if "the proper null hypotheses are specified," arguing that the most common mistake is testing against a null of zero when other null is of interest. However, in transportation, it is common that null hypotheses are implied based on the objectives of the study and are not stated explicitly. For example, Khan, Kockelman and Xiong (*4*) state as their objective "to evaluate the effects of built environment variables on the use of non-motorized travel modes." As such, a priori hypotheses of parameter sizes, which are not directly inteptable, are very rarely specified, with the clear expection of theoretically defined parameters such as the scale parameter in the nested logit model. As such, such as question would not be very informative and hence removed.
  MZ-5 on whether coefficients are carefully interpreted, was removed since it largely overlaps with Q2 and Q8. MZ-7 on using statistical significance as the only criteria of importance at its first use, and MZ-15 on avoiding using statistical significance as the only important criteria after the "crescendo" were removed because such issues could be covered to a large extent by Q7~Q10. Finally, MZ-16 on whether statistical significance was decisive, and conveyed the sense of an ending, was deleted because it was rather ambiguous, hard to operationalize, and its content could also be covered with Q7~Q10.

## 4. MAIN FINDINGS
The results of the review are summarized in **Table 1**. The key findings are listed below, with the values in parenthesis summarizing the values reported by McCloskey and Ziliak (*2*) for reference purposes.
1.  67% (MZ:70%) of reviewed studies did not distinguish statistical significance from economic, policy or scientific significance.
2.  86% (MZ: 72%) of studies did not discuss the scientific conversation within which the magnitude of a coefficient can be judged to be "large" or "small."
3.  62% (MZ: 59%) of studies ambiguously used the word "significant" to mean statistically different from the null sometimes and to mean practically important at other times.
4.  39% (MZ: 53%) explained model results exclusively based on the sign of the coefficient.
5.  24% (MZ: 32%) explicitly stated to have used statistical significance as an exclusive criterion to drop variables from a model.
6.  0% (MZ: 4%) of the reviewed studies considered the statistical power of the tests.

*Parady and Axhausen*

7. 0% of the reviewed studies reported confidence intervals (of effect magnitudes) and used them to interpret economic or policy significance. 7% did however report these intervals but did not explicitly use them in the discussion.

**TABLE 1 Answer to the 15 questions in the transportation field**

| Does the article... | Applicable articles | % Yes | Out of which: | |
| --- | --- | --- | --- | --- |
| | | | Comprehensively /largely | Limitedly /partially |
| Q4: Consider the power of the test? | 94 | 0.00 | - | - |
| Q5: Examine the power function? | 0 | - | - | - |
| Q15: Report effect confidence intervals, using them to interpret economic significance not merely as a replacement for pointwise statistical significance? | 95 | 7.37 | 0 | 7.37 |
| Q10: Discuss the scientific conversation within which an effect or other quantity of interest would be judged large or small? | 95 | 13.68 | - | - |
| Q12: Do a simulation to determine whether the estimated effects or other quantities of interest are reasonable and/or to better illustrate the magnitude of estimated effects? | 95 | 29.47 | - | - |
| Q13: In the conclusions and implications sections, keep statistical significance separate from economic policy and scientific significance? | 95 | 32.63 | - | - |
| Q9: Make a judgement on effect magnitudes? | 95 | 36.84 | 13.68 | 23.16 |
| Q14: In the estimation, conclusions, and implication sections, avoid using the word "significance" in ambiguous ways? | 93 | 37.63 | - | - |
| Q7: In the model results section, eschew "sign econometrics"? | 94 | 60.64 | 27.66 | 32.98 |
| Q8: Discuss the magnitude of estimated effects or other quantities of interest? | 95 | 64.21 | 33.68 | 30.53 |
| Q2: Use coefficients to calculate elasticities, or some other quantity that addresses the question of "how large is large"? | 95 | 65.26 | 45.26 | 20.00 |
| Q11: Avoid choosing variables for inclusion solely on the basis of statistical significance? | 94 | 75.53 | - | - |
| Q3: Report all traditionally reported statistics? | 95 | 76.84 | - | - |
| Q1: Report descriptive statistics for model variables? | 95 | 78.95 | 65.26 | 13.68 |
| Q6: Eschew "asterisk econometrics"? | 94 | 100.00 | - | - |

For Q14, paper that do not mention the word "significant" are marked as NA.

">8

For Q4,Q6, Q7 and Q11 papers that did not directly show model results (i.e., reported in a separate article) are marked as NA.

**4.1 Size matters: on the reporting and discussing of effect magnitude**
Focusing specifically on the questions concerning effect size, first we want to highlight two very positive findings. First is the complete eschewing of "asterisk econometrics," that is, the ranking of coefficients according to the absolute size of the test statistic. The second, is the high levels of reporting of descriptive statistics. 79% (MZ: 31%) of studies did so, 65% did so extensively, and 14% did so only partially. As we discussed earlier, such information is necessary for properly interpreting model results. 77% of studies also reported all traditionally reported model statistics, that is, coefficients, measures of significance, and goodness of fit. However, while we do not oppose such reporting per se, we want to underscore this should not be the highpoint of the paper, as it often is. If anything, it should be supplementary material the reader can consult if required, thus moving the paper towards matters of practical importance. In that regard, 65% (MZ: 67%) of studies used coefficients to calculate elasticities, or some other quantity of interest that conveys magnitude. 45% did so extensively (either for the majority of the model variables, or the variables the authors explicitly stated being key to the study) while 20% did so only partially. In line with this, 64% (MZ: 80.2%) of studies explicitly discussed the magnitude of estimated effects or other quantities of interest. 34% did so extensively, 31% did so only partially (difference due to rounding error). That is, in addition to reporting magnitudes, the authors make the decision to discuss the model results in terms of such magnitudes. For example, exploring the impacts of walk and bike infrastructure on mode choice, Aziz et al. (*5*) report in a very orthodox manner that "*the direct elasticity value indicates that 1% increase in the total bike lane proportion (normalized by area) in the home and work census tracts will increase the probability to choose bike by 1.13%.*" Heinen & Ogilvie (*6*) also discuss effect magnitudes in the context of the impacts of the introduction of a new guided busway in Cambridge, UK, stating that "*the results correspond with individuals living, for example, 4 km from the busway being from 60% to 70% more likely (depending on the indicator of variability) to have increased their active travel share [more than 20%] than those living 9 km away.*"

It is important to note that among papers classified as "yes, partially" for Q2, a large share simply reported one or several measures of value of travel time, and then reverted back to reporting only coefficient signs for all the other variables. In fact, 39% (MZ: 53%) of studies explained model results exclusively based on the sign of the coefficient, with no reference to effect magnitudes or other quantities of interest. While convention plays a role here, it is worth noting that some authors did in fact, after reporting models exclusively sign-wise, go on to conduct simulations to determine whether the estimated effects or quantities of interest are reasonable or to evaluate policy effects. That being said, 22% of papers did not report or discuss any magnitude whatsoever. That is, the discussion was exclusively limited to sign direction and statistical significance.

Regarding the question of "how large is large?" it is worth pointing that 63% of studies failed to make a clear judgement of magnitude. We make the distinction here from the question regarding explicitly discussing estimated magnitudes (Q8) because such magnitudes were frequently discussed in relative terms, and in many cases the authors fell short of a judgement on whether the quantity of interest was "large" or "small" or "policy-relevant" or "not relevant" in absolute terms based on any criteria. In this regard, Khan, Kockelman and Xiong (*4*) make clear judgements of magnitude when they state that "*network connectivity (measured as 4-way intersections within 0.5 mile) plays a major role: a single standard deviation change in this variable is estimated to increase walking probability by 34%*" and go on to state that "*parking prices and free-parking availability variables were not found to have much of an effect.*" It is clear from these statements that based on the authors judgement, network connectivity is a variable of practical importance. Here we also want to highlight that they evaluated magnitudes by

<em>Parady and Axhausen</em>estimating percentage change in dependent variables given one standard-deviation changes in the independent variables. Such an approach is designed to overcome one key limitation of traditionally reported elasticities (defined as the percentage change in the dependent variable given one-percent change in the independent variable,) which is that a 1% change might be easier to achieve for some variables than others.

de Luca and Di Pace (*7*) also make clear judgments of magnitude when they discuss the magnitude of value of travel time estimates and state that "*aside from being similar to those estimated in different Italian case studies* (*8*)*, [the magnitude] indicates the extreme importance of parking location. Assuming that the average one-way travel monetary cost is equal to 3 €, 10 min walking time (about 700 m at 4 km/h) is more than half of the whole travel monetary cost.*" This is an ideal form, as it gives a clear economic interpretation of the quantity in question and a clear judgment of its magnitude. In addition, they compare estimated magnitudes to similar studies in the literature.

The questions of "how large is large?" is, however, a difficult question with no easy immediate answer, if anything, underscoring the importance of addressing it. The very concepts of "small" or "large" are difficult to characterize and might require some degree of convention. In his seminal work on power analysis, Jacob Cohen (*9*) argued that *"all conventions are arbitrary, one can only demand they not be unreasonable."* And while he noted that it was desirable to have, and actually proposed and characterized universal effect size measures, free of unit variability and applicable to various research issues and stastitical models, he warned that *"the meaning of any effect size is, in the final analysis, a function of the context in which is embedded."* Thus, addressing the question of how large is large requires a clear understanding of the scientific context of the study. On this point, 86% (MZ: 72%) of studies failed to discuss the scientific conversation within which the magnitude of an effect or other quantity of interest can be judged to be "large" or "small" by referencing to values reported in the literature of at least one variable. Allard and Moura (*10*) provide this scientific context by reporting a table comparing several values of time and willingness to pay for long distance intermodal service characteristics, the object of their study.

For some variables, judgement of magnitude is not that straightforward, and even impossible to discuss in economic terms. This is especially so for latent constructs, where the meaning of unit changes or percentage changes are not clear-cut. Hess et al. (*11*) address this issue in the context of latent attitude constructs, and propose that instead of arbitrarily looking at percentage changes in attitudes, pointless due to the scale of such constructs, it would be more meaningful to test what would happen if everyone's attitudes were like those of a particular segment of the population.

### 4.2 The issue of power
Another issue worth highlighting is that none of the reviewed studies considered the statistical power of the tests (MZ: 4%). While a comprehensive exposition of statistical power is beyond the scope of this paper, we will briefly discuss the main issues with the goal of sparking a well-deserved debate on the matter.

The statistical power of a test gives the probability that the test will *correctly* reject the null hypothesis when the null is *actually* false, that is, the probability of avoiding Type II errors (false negatives). Power is a function of sample size, statistical significance and more importantly, effect size. It is most commonly used to test the power a statistical test had on a completed study, and to calculate necessary sample sizes given anticipated effect sizes and power (*9*). In other words, it is used to answer two questions: (a) assuming that the effect we are looking for actually exists and has magnitude *m*, for sample size *n*, what is the probability we will detect such effect (i.e., correctly reject the null) at significance level α? And (b) what sample size do we need in our study to identify an effect of magnitude *m*, at significance level α with a given power level?

To give a concrete example, a test with power of 0.2, means that the researchers will mistakenly accept the null hypothesis four out of five times. In the words of Ziliak and McCloskey (*3*), *"power puts a*

*Parady and Axhausen*

*check on the naïveté of the gullible."* This is particularly critical for "small" effects which will require larger samples to be detected.

Specifically related to multivariate modeling, where multiple tests are conducted, Maxwell (*12*) showed in his analysis of underpowered studies in psychology, that required samples sizes to detect at any given power level of (a) any single prespecified effect, (b) at least one effect, or (c) all effects, differ. For example, he showed that for a multivariate linear regression with five predictors (correlation between each predictor with other predictors and with the outcome variable = 0.3, n=400 and $α=0.05$) while the power of correctly detecting at least one effect (that is, correctly rejecting the null when it is actually false) was >.99, the power for correctly detecting all effects was only .22. Given that review studies continue to show lack of power in the literature, he continued, tests of indiviudal hypotheses often lack sufficient power, even when adequate power exists for detecting an effect somewhere in the collection of tests. In that regard, while in the transportation field, at the usual sample sizes of large-scale household travel surveys, power issues might not be much of a problem, in the studies we reviewed, the median sample size (when several samples are reported, the minimum was used) was 1,404 (choice situations) and the $20^{th}$ percentile was 527. Although we do not make claims of the applicability of Maxwell's findings to discrete choice models, his results do underscore the need to address such concerns in the transportation field. While many power studies have been conducted in fields like psychology (*13*, *14*) and education (*15*), to the best of our knowledge, such analyses have not been conducted in the transportation field, so the state of affairs is not known. It is important, however, to note that the research in the psychology field rests on the extensive work on statistical power by Cohen (*9*, *13*) who defined "scale-free" conventions to characterize "small," "medium," and "large" effects, and while Cohen largely focused on t-test for means, correlation coefficients, proportional differences, and linear regression, the literature on statistical power for discrete choice models is scarce (*16*, *17*). Finally, regarding sample size determination, it must be pointed out that while there is a comprehensive literature dealing with sample size for discrete choice experiments (*18*, *19*) existing theory largely ignores the issue of minimum sample size requirements in terms of power (*17*) .

**4.3 Conflating statistical significance with practical importance**
In the background of the magnitude discussion is the misguided focus on statistical significance and its frequent conflation with practical importance. This is evident in the abovestated fact that 22% of papers did not report or discuss any magnitude whatsoever. This means that one in five reviewed papers completely defined the importance of their findings based on statistical significance. 24% (MZ: 32%) of studies explicitly dropped variables solely based on statistical significance, and 62% (MZ: 59%) of studies used the term "significant" either conflating it with with practical importance or in a way the reader cannot discern which meaning the author is pointing to. In some cases, the interpretation of the size of the t-statistic was misinterpreted as a measure of effect size. In a study of social interaction effects on the decision-making process, Kamargianni et al. (*20*) state of a latent construct of walking preference that *"this component is the most statistically significant variable…indicating the strong influence that parents have on the development of their children's attitudes towards walking"* misinterpreting a large t-statistic with a strong influence on outcome. Similarly, Qin et al. (*21*) argue in a study of mode-shifting behavior that *"bus service level has the most significant positive t-value, which indicates that improving the bus service level can increase the shifting proportion of car travelers to bus significantly."* Here it is also unclear whether or not the "significant" increase in modal shift proportion is meant to signify a considerable increase in practical terms, or just a difference statistically different from zero.

Finally, 67% of studies mixed statistical significance and practical importance in the discussion and conclusion sections, the most common practice being reporting a set of variables as important, based on statistical significance, without having properly discussed these in terms of their practical importance, that is, their effect magnitudes.

*Parady and Axhausen***4.4 Do top journals do any better?**
To answer this question, we developed a very simple score to get an idea of the overall performance of the reviewed papers. We scored each "Yes, comprehensively/largely" as 1, each "Yes, limitedly/partially" as 0.5 and each "No" as 0, averaged it over the total number of valid questions for each paper and normalized to 100. Q3 was excluded from this score as we do not believe that mechanically reporting all traditionally reported statistics necesarily implies good practice.

After calculating the mean and standard deviation of scores for articles in top journals (defined as first quartile journals in the Scimago Journal Ranking in the field of transporation) we found that articles in top journals performed marginally better (N: 69, mean: 44.4, sd: 40) than non-top journals (N: 26, mean: 42.6, sd:39.3) but the difference was surpisingly small, a difference that was, for what is worth, not statiscally significant.

## 5 RECOMMENDATIONS TO THE FIELD
Based on the discussion above, we put forth a set of reccomendations aimed at shifting the focus away from statistical significance towards proper and comprehensive assessment of effect magnitudes and policy relevant matters:

*Make reporting of effect magnitudes and their confidence intervals mandatory.*
Statistical significance should not be more than one of many criteria of evaluation, but it should certainly not be the most important one. The discussion of statistical models should focus on effect magnitude and other policy relevant quantities. In that regard, confidence intervals of effect sizes (not of coefficients) give a clearer image of effect magnitude and the levels of uncertainty surrounding the estimates, and does without what Maxwell (*12*) calls that "*air of finality*" that the presence or not of asterisks tend to convey. For elasticities and marginal effects, confidence intervals are usually estimated via bootstrapping (*22*).

Note that while we do not oppose reporting model coefficients, due to their lack of direct interpretability, these should be relegated to secondary position in the paper, or even an appendix.

*Provide to the extent possible judgements of magnitude that convey what the authors consider are "small," "medium," or "large" effects (or other quantities of interest) and the basis for such judgement.*
While we acknowledge this is certainly not an easy task, there is a discussion to be had regarding what effects or quantities are policy relevant and how to assess such relevance. Furthermore, such discussions should ideally be accompanied by a discussion on the cost implications of changing the policy variables in question. As discussed earlier in the context of elasticities reporting, while the effect of a 1% increase in a policy variable might be practically larger for some variables than others, the costs associated with that 1% increase might be higher as well. While the approach adopted by Khan, Kockelman and Xiong (*4*) accounts for such differences to some extent, an explicit discussion of the cost implications of such increases, while rarely conducted, is of high importance to policy making and should be actively encouraged.

*Compare, whenever possible, effect magnitudes or other quantities of interest to existing studies.*
For the most regularily reported values, such as value of travel time, there is a miryad of studies reporting such values for many contexts (*23–25*), so there are no reasons why such comparisons cannot be made. For less often reported values, given the irregularities in reporting discussed above and differences in variable definition and measurement there will be certainly times when such a task will be difficult, but should magnitude reporting become mandatory and authors strive to provide judgements on such magnitude, in time, proper discussion of scientific context should be widespread, thus catalyzing a virtuous cycle of proper reporting practices and discussion of magnitudes.

*For new studies, take statistical power into consideration when defining sample size to guarantee the effects the research wants to detect can in fact be detected with enough power. For studies using secondary data (i.e., national household survey data, etc.) report post-hoc power levels of tests reported in the study.*

Certainly, the literature on this issue is rather scarce, but the work of deBekker-Grob et al. (*17*) should be a starting point.

## 6   CONCLUSION
In this study we reviewed the academic transportation literature published between 2014 and 2018 to evaluate where the field stands regarding the use and misuse of statistical significance in empirical analysis, with a focus on discrete choice models. Our results showed repeated errors in the use of statistical significance and a lack of a clear focus on effect magnitudes for a considerable share of studies. We want to reiterate that the ultimate objective of transportation academic research is to better inform transportation policy design and implementation, which requires proper discussion of effect magnitudes and their practical implications. In that sense, note that the purpose of this study is not to criticize the field but to point out ways it can realign itself better with this ultimate objective.

## 7   ACKNOWLEDGMENTS
This work was supported by JSPS KAKENHI Grants Number 20H02266.

## 8   AUTHOR CONTRIBUTIONS
The authors confirm contribution to the paper as follows: study conception: Giancarlos Parady, Kay W. Axhausen, study design, literature review & draft writing: Giancarlos Parady, draft revision: Giancarlos Parady, Kay W. Axhausen